\documentstyle[revtex]{aps}            

\begin{document}
\draft

\begin{title}
Spin liquid phase for the  Frustrated Quantum  Heisenberg\\
Antiferromagnet  on a square lattice
\end{title}

\author{Jaime Ferrer}
\begin{instit}
Serin Physics Laboratory, Rutgers University, P.O. Box 849,
Piscataway, NJ 08855-0849
\end{instit}
\receipt{April 1992}

\vspace{0.8cm}                           
\begin{abstract}
The existence of a spin disordered ground state
for the frustrated $J_{1}-J_{2}-J_{3}\,\,$
Quantum Heisenberg
Antiferromagnet on a square lattice is reconsidered.
It is argued that there is a unique action which is continuous
through the whole phase diagram, except at the Lifshitz point,
so that the Neel and helicoidal  states
can not coexist and that there has to be
an intermediate spin liquid state. To show it,
a detailed study
combining Spin-Wave
Theory, Schwinger Bosons Mean Field Theory and a
scaling analysis of the appropriate hydrodynamic
action is performed.
When done carefully, all these theories
agree and strongly
support the existence of the spin liquid state.
\end{abstract}
\vspace{0.7cm}                           


\narrowtext

\section{Introduction}
The past few years have seen a considerable effort in understanding
the nature of the ground state and excitations of low-dimensional
Quantum Antiferromagnets, due to its close link to high-$T_{c}$
superconductors \cite{anderson1,chn,shirane,lro}.
One of the most fundamental discussed issues is the existence and
properties of spin liquid states.
\cite{bethe,haldane,affleck}.

The way in which one can induce a spin disordered state
in a magnet is by increasing  fluctuations about the ordered
state, for instance,
by adding frustration to the bare Heisenberg model
--antiferromagnetic coupling constants to next to nearest neighbors--
\cite{chandra}.
We study in this paper the phase diagram of
the quantum $J_{1}-J_{2}-J_{3}\,\,$ model at
zero temperature (all $J_{i}\ge 0$).
We show in figure 1
the phase diagram of the corresponding classical model.
\cite{moreo}.
We see that there are four possible states --Neel, collinear and two
helicoidal-- separated by boundaries. The transition between Neel and
collinear states is discontinuous because they have different
symmetry.
The same happens in the boundary between the $(q,q)$
and $(q,\pi)$ helicoidal states. The other transitions (marked with
dashed lines) are continuous and separate states of the same symmetry.
The {\em Classical Critical Line (CCL)} $J_{1}-2J_{2}-4J_{3}=0$
is the separatrix between Neel and helicoidal states,
and its endpoint, $J_{3}=0\,\,$, is a Lifshitz point.

The first studies on the phase diagram  of
the quantum model were performed by Chandra and
Doucot \cite{chandra} (for $J_{3}=0\,\,$)
using Linear Spin Wave Theory (LSWT)
and by Ioffe and Larkin \cite{ioffe}, who made a Renormalization Group
(RG) analysis of a generalized nonlinear $\sigma$ model.
Both of them
concluded that the combined effect of quantum fluctuations and
frustration
were strong enough to destroy long-range-order (LRO) along some
critical line yielding a spin liquid state. Later on, Moreo et al
\cite{moreo} and  Chubukov \cite{chubukov}
generalized the LSWT analysis to $J_{3}\neq0$ (see figure
1). This scenario
was supported by series expansion \cite{series}, finite-size
diagonalizations \cite{dagotto,schulz} (for $J_{3}=0\,\,$)
and a momentum-shell RG study \cite{einarson}.
On the other hand, several mean field-BCS-type theories of interacting
bosons have challenged this result, either for $J_{3}=0\,\,$
\cite{starykh,xu,mila,bergomi} or for the general case \cite{read}.
They predict that the stability of the Neel state is enhanced over
the classical case due to quantum fluctuations (Villain's {\em order
from disorder} \cite{villain}) along the whole critical line, while
that of the spiral state is not modified in an essential way.

The purpose of this paper is to clarify this controversy. Our main
points will be:
\begin{enumerate}
\item The {\em CCL} separates states with the same symmetry. The
transition is continuous classically, in the sense that the parameters
of the long wavelength action ${\cal S}[J_{i}]$ vary smoothly along
the
transition. We argue that the same happens for the quantum case,
${\cal S}[J_{i},S]$. In particular, we show that, although there is an
enhancement of the stability of the Neel state close to the {\em CCL},
there is a bigger decrease of the stability of the spiral state.
This means that the {\em CCL} is continued in a {\em Quantum Critical
Plane, QCP}, which depends on S and is tilted upwards with respect to
the plane $J_{1}-2J_{2}-4J_{3}$  (see figure 1).
\item Because the quantum action is continuous through the {\em QCP},
we can use a generalized Sigma model to describe the physics of the
transition along the {\em QCP} and at its Neel side.
We evaluate that action along the {\em QCP} using a
Renormalization Group analysis and find that the
relevant coupling constant always flows to strong coupling meaning
that the spin state is disordered. We then proceed to show that the
spin liquid state exists not only along that plane but also in a
finite region around it.
As a corollary, we link this scaling analysis for the strongly
frustrated magnet with the one performed in reference 2, for the
weakly frustrated case.
\item The end point of the {\em CCL} is a Lifshitz point. In this
case, the counterpart of the Neel state is not spiral, but  collinear.
Because both states have different symmetry, the quantum action for
both states is different, and therefore we cannot establish any
continuity principle. We discuss this case in the last section.
\end{enumerate}

\section{Microscopic study of the $J_{1}-J_{2}-J_{3}\,\,$ model}
The purpose of
this section is to study the stability diagram of the quantum
$J_{1}-J_{2}-J_{3}\,\,$ model around the {\em CCL}. Because
the whole line has the same physics,
we can concentrate our analysis just to one,
representative point of the  phase diagram. We
will choose the plane $J_{2}=0$ for simplicity, in which case the {\em
CCL} is $J_{3}=0.25 J_{1}$. We will work in units in which $J_{1}=1$.

The energy of the classical $J_{1}-J_{2}-J_{3}\,\,$ model is given by
\begin{equation}
E = \sum_{<i,j>} J_{ij} \vec{S}_{i}\cdot\vec{S}_{j}=\frac{S^{2}}
{2}J_{Q_{0,cl}}
\end{equation}
where $J_{Q}$ is
the Fourier transform of $J_{ij}$ and $\vec{Q}_{0,cl}$ is the
pitch wave vector of the spiral state, which is obtained by
minimizing the energy: $\partial_{Q} J_{Q_{0,cl}}=0$. $\vec{Q}_{0,cl}$
is a continuous
function of $J_{2}/J_{1}$ and $J_{3}/J_{1}$, with a jump in its
derivative at the critical value $J_{1}-2J_{2}-4J_{3}=0$. This
equation
defines the {\em Classical Critical Line}. The
classical spin stiffness is given by
\begin{equation}
\rho^{z}_{ij}=\frac{\partial^{2}E}
{\partial Q_{i}\partial Q_{j}}|_{\vec{Q}_{0,cl}}
\end{equation}
and it also varies continually with $\vec{Q}_{0,cl}$.

We can then write the
most general action for an helical magnet by noting that the order
parameter space is $O(3)\times U(1)/U(1)$ \cite{azaria,dombre0,piers}
\begin{equation}
{\cal S}= -\frac{1}{2}\int d^{2}x Tr\{A_{i}\,P_{i}\,A_{i}\}=
-\frac{1}{2}\int d^{2}x \rho_{a} (A_{i}^{a})^{2}
\end{equation}
where $A_{i}=A_{i}^{a}\,T_{a}= g^{-1}\partial_{i}g$ is a pure gauge
field,
$g(x)\,\epsilon \,SO(3)$ and $T_{a}\,\epsilon$ Lie$[SO(3)]$. The gauge
field is equivalent to a twist of the order parameter and serves to
define the spin stiffnesses, $\rho_{a}=(\rho_{x},\rho_{x},\rho_{z})$.
The action $S[\rho(\vec{Q}_{0,cl})]$ is unique for both Neel and
helical
states and the spin stiffnesses are continuous throughout the whole
phase diagram.

We argue that, after introducing quantum fluctuations, there is still
a unique action for the whole phase diagram, which varies continually
when passing from the Neel to the spiral state through the {\em
Quantum Critical Plane}.
The action
picks up an extra piece due to the fact that the fields become
dynamical and an implicit dependence on S through $\vec{Q}_{0}(S)$:
\begin{equation}
{\cal S}=
-\frac{1}{2}\int dx_{0}d^{2}x (\chi_{a} (A_{0}^{a})^{2}+\rho_{a}
(A_{i}^{a})^{2})
\end{equation}

All this implies that if the stability of the Neel state gets
enhanced due to quantum fluctuations, (1) the pitch wave vector of the
helical state also renormalizes so as to adjust smoothly to the new
boundary of the Neel state, where its value is $(\pi,\pi)$, (2) the
spin stiffness of the helical phase is equally decreased, so that both
go to zero at the same point. To substantiate this ideas, we will
perform in this section a Spin Wave Theory (SWT) study of
the magnetization, the
pitch wave vector and the spin stiffness and complement it, eventually
with results from Schwinger Bosons Mean Field Theory (SBMFT). Details
of the derivation and notation are relegated to appendices A and B.

As a starting point,
let's determine the phase diagram from LSWT and SBMFT.
The magnetization for the helicoidal $\vec{Q_{0}}=(q_{0},q_{0})$ and
Neel $\vec{Q_{0}}=(\pi,\pi)$ states from LSWT is
\begin{equation}
<S^{z}>=
=S+\frac{1}{2}-\frac{1}{2N}\sum_{k}\frac{J_{k}+J_{\pm}
-2J_{Q_{0}}}{E(k)}
\end{equation}
$\vec{Q_{0}}$ is obtained by minimizing the ground state energy:
$\cos(q_{0})\sim \cos(q_{0,cl})=-
\frac{1}{4J_{3}}$ (for the Neel state, $\cos(Q_{0})=-1$). The
denominator is the energy of the spin waves. The value of S for which
the staggered magnetization $<S^{z}>$ is zero determines the
boundaries of the long range ordered states.

We plot in figure 2 the stability diagram we have obtained using
LSWT and SBMFT.
The theories are in disagreement: (1)LSWT predicts a decrease of the
stability of
both ordered states, with a line of second order phase transitions
ending at the critical point; (2) SBMFT  predicts an enhancement of the
stability of the Neel state. This enhancement is also seen in
numerical diagonalization studies \cite{moreo}. On the other hand, the
boundary of the spiral state doesn't move towards the right as we
introduce quantum fluctuations. This is a spurious result, due to the
fact that large-N theories don't give the correct Goldstone mode
structure for the spin wave spectrum. As a result, we obtain a first
order phase transition.

Let's concentrate for a moment on the Neel state and compute
the expression for the staggered magnetization close to
the classical
frustrated point. Evaluating the leading divergences we obtain
\begin{equation}
<S^{z}>\simeq S+\frac{1}{2}-\alpha \ln(\rho_{cl})+\frac{\beta}{S}
\frac{1}{\rho_{cl}}+O(\frac{1}{S^{2}})
\end{equation}
where $\rho_{cl}=J_{1}-4J_{3}$ is the classical  spin stiffness
which we use here as a cutoff for the infrared divergent integrals.
This series can be resumed: computing ${\cal E}(k)$ --the spin wave
energy to next to LSWT order in $1/S$--, we find:
\begin{eqnarray}
<S^{z}>&\simeq& S+\frac{1}{2}-\gamma\ln(\rho(S))\\\nonumber
\rho(S)
&\simeq&\rho_{cl}
-\frac{4J_{3}}{S}(\alpha_{3}-\alpha_{1})
\end{eqnarray}
which is the LSWT result with a renormalized spin stiffness. It
serves us to define the {\em QCP},
i.e.: the line where the spin stiffness, computed to next to
LSWT order  is zero. As shown  in figure 2, $\rho(S)$ is tilted
to the right of the classical value, with a slope $\sim 5.2$.
This fact supports the {\em order from disorder}
conjecture from the point of view of SWT.

Based in our symmetry argument, we argue that this line has to be the
same for both the helicoidal and the Neel states (at least
for large S).
In other words, the enhancement of the stability of the Neel
state is accompanied by a similar reduction of the stability of the
helicoidal one
so that both states match each other continuously.
To prove it, we
compute the LSWT correction to the pitch wave vector and to the
stiffness:
\widetext
\begin{eqnarray}
q_{0}^{i} - q_{0,cl}^{i}&=&\Delta q_{0}^{i}=
\frac{1}{2 S \partial^{(2)}_{i}J_{Q_{0,cl}}}\sum_{k}\partial_{i}
J_{\pm}
\left(\frac{J_{k}-J_{Q_{0,cl}}}{J_{\pm}-J_{Q_{0,cl}}}\right)^{1/2}
+O(\frac{1}{S^{2}})\\\nonumber
\rho_{ii}^{z}&=&\frac{S^{2}\partial^{(2)}_{i}
J_{Q_{0,cl}}}{2}\\\nonumber
&&+\frac{S}{2}\left\{
\partial^{(2)}_{i}J_{Q_{0,cl}}-(\sin(q_{0,cl})+8
J_{3}\sin(2q_{0,cl}))S\Delta q_{0}^{i}\right.\\\nonumber&&\left.
+\frac{1}{2}\sum_{k}\frac{(2J_{Q_{0,cl}}-J_{k}-J_{\pm})
\partial^{(2)}_{i}J_{Q_{0,cl}}-(J_{Q_{0,cl}}-J_{k})
\partial^{(2)}_{i}J_{\pm}}{E(k)}\right\}+O(1)
\end{eqnarray}
\narrowtext

The spin stiffness of the unfrustrated magnet is
$\rho=\rho_{cl}(1-\frac{0.118}{S})$ ($=0.764\rho_{cl}$ for spin 1/2).
This estimate is in
accordance with that obtained using the renormalization of the spin
wave velocity and the susceptibility \cite{chn},
$\rho=\rho_{cl}Z_{\chi}Z^{2}_{c}\simeq\rho_{cl}(1-\frac{0.276}{S})
(1+2\times\frac{0.079}{S})$.

We draw in
figure 3(a) the LSWT correction to $q_{cl}$ for the spiral state
to show that it matches with $(\pi,\pi)$ continually, although the
boundary of stability of the Neel state be shifted to the right.
Figure
3(b) is a plot of the LSWT correction to the spin stiffness, and
shows that it is enhanced for the Neel state close to the critical
point, signaling the {\em order from disorder phenomenon}.
The important point is that
it is decreased by a bigger amount on the other side.
We stress that this
beautiful results are already obtained in LSWT.

Before entering in the next section, we would like to summarize the
most important conclusions we have obtained so far. We can very safely
state that the stability of the Neel state is enhanced over the
classical values over the whole critical line,
and that
this is due to the fact that the spin stiffness $\rho(S)$ gets
renormalized
upwards due to quantum fluctuations, so that it defines a
{\em QCP} tilted to the right of the {\em CCL}.
The stability of the helicoidal state is
reduced in the same way, so that both states match smoothly and the
long wavelength physics is continuous through the transition point.
Because of that, a modified $\sigma$ model which describes the spin
wave excitations of the strongly frustrated Neel state is appropriate
to study the {\em QCP} and its neighborhood. These
conclusions don't apply to
the $J_{1}-J_{2}\,\,$ model due to the fact that in that case
the two states have different symmetry and the spin stiffnesses close
to the tricritical point are different.

\section{Scaling Analysis}
We devote this last section to perform a Renormalization Group study
of the hydrodynamic
action appropriate for the strongly frustrated magnet.
It is a generalization
of the $O(3)/U(1)=S^{2}$ quantum non linear $\sigma$
model. We will review in section 4.1 well-known
results for the
conventional $O(3)$ case, published by other authors, which
are pertinent
to our discussion. In section 4.2, we pass to the strongly
frustrated case, and  show that along the {\em QCP}
the system always flows towards strong coupling. The results of
this section complement our previous findings from microscopic models
and give us a very consistent
picture of the phase diagram of the $J_{1}-J_{2}-J_{3}\,\,$
model.
\subsection{Unfrustrated case}
It is generally believed that the long-wavelength action of a
non-frustrated Quantum Antiferromagnet
in d dimensions is given by a $(d+1)$-dimensional non-linear
$\sigma$ model \cite{haldane1}
\begin{eqnarray}
{\cal S}&=&
\frac{\rho^{0}}{2}\int_{-\infty}^{\infty}dt\int_{a}d^{d}r
\left\{(\nabla\hat{n})^{2}
-\frac{1}{c_{0}^{2}}(\partial_{t}\hat{n})^{2}
\right\}\\\nonumber
&=&\frac{1}{2g_{0}}
\int_{-\infty}^{\infty}d\tilde{t}\int_{1}d^{d}\tilde{r}
\left\{(\nabla\hat{n})^{2}-(\partial_{\tilde{t}}\hat{n})^{2}\right\}
\end{eqnarray}
where $\tilde{r}=r/a$, $\tilde{t}=t\,c/a$ are dimensionless variables,
$\rho_{0}=J_{1}S^{2}a^{2-d}$ is the bare  spin stiffness,
$c_{0}=2\sqrt{d}J_{1}Sa$ is the spin wave velocity and
\begin{displaymath}
g_{0}=\frac{c_{0}}{\rho_{0}a^{d-1}}
=\frac{1}{\sqrt{\rho_{0}\chi_{0}a^{d-1}}}=\frac{2\sqrt{d}}{S}
\end{displaymath}
is the bare coupling
constant. All these magnitudes are defined at the scale
of the lattice cutoff, a.

In 1+1 dimensions,
one has to add to this action a topological $\theta$
term, which distinguishes between integer and half-integer spins
\cite{haldane1}. It has been proven that such a term gives a
zero contribution in 2+1 dimensions for any topological
configuration of the n-fields
\cite{haldane2,dombre,einarson2}.
Moreover, as long as the n-field is continuous at all points no Hopf
term appears in the action. Once this condition is relaxed, though,
that term appears and affects the physics of the system
\cite{haldane2,read2}. Such a situation could arise if the $\sigma$
model  were in its disordered phase.

One of the main issues
concerning the nonlinear quantum $\sigma$ model is the
connection with the Classical one, which is much better known. This
connection can be done
in principle when the contribution of the Berry phase is
zero.
Strictly speaking, the mapping is rigorous only for large S and long
wavelengths. We therefore face two problems when we try to link it
with the microscopic quantum models of interest:
\begin{itemize}
\item We don't know by
certain if the action is still valid in the extreme
quantum limit. In this respect, nevertheless, we can argue that it
gives the correct spin wave spectrum
and interaction
between spin waves at long wavelengths, as one can check by
calculating the staggered magnetization, spin wave energy and
damping,\ldots in the first few orders of perturbation theory
\item A more difficult point is the identification of the parameters
entering in the action in terms of the ones appearing in the
microscopic model.
The problem is that the values obtained in the derivation are well
suited only for large spin and long wavelength scales, and they don't
have to hold necessarily for small S. In fact, they don't. One
possibility  is to connect those values with the ones obtained by SWT.
But it is not clear
which order in $1/S$ should one use as input of the continuum model.
That should depend on how strongly the Goldstone modes interact and,
therefore, on the dimension.
\end{itemize}

Polyakov pioneered the work on the Classical NL$\sigma$M by
recognizing that close to its lower critical dimension Goldstone modes
interact strongly and become critical. He proceed then to solve the
problem using a Renormalization Group  approach.
Later on, the same problem was addressed by Br\'{e}zin and Zinn-Justin
\cite{brezin} in the context of the $\epsilon$-expansion
($\epsilon=d-2$) to order
$\epsilon^{2}$ and by Nelson and Pelcovits
\cite{nelson} using
momentum-shell recursion relations \cite{wilson} to one
loop order, i.e.: to first order in $\epsilon$. The approximation is
expected to be very good in two dimensions, but it is not so clear
that this is the case in three (where $\epsilon=1$).

Once we have an action to describe the system in 2+1,we have to
distinguish
between short wavelength quantum critical fluctuations and
the long wavelength antiferromagnetic magnons. The crossover length
between both is the Josephson correlation length $\xi_{J}$
\cite{amit}. $\xi_{J}$ tells us the extent to which the quantum
model can be mapped to the classical one. This problem was addressed
by Chakravarty, Halperin and Nelson in an important publication
\cite{chn}.
CHN combined the techniques of Nelson and Pelcovits with
the method developed by Hertz \cite{hertz} to deal with quantum
critical phenomena at finite temperatures in
two spatial dimensions. They showed that:
\begin{itemize}
\item At $T=0$ the model is equivalent to a 3-dimensional classical
one where $g\sim 1/S$
plays the role of the temperature. In that case, the QNL$\sigma$M
has a phase
transition from a Neel ordered to a (quantum) disordered state, at
a value $g_{c}$ of the coupling constant. The Josephson correlation
length turns out to be of the order of a few lattice spacings for
$g<g_{c}$, but grows with g until it diverges at $g_{c}$.
\item For $g>g_{c}$ the system is in a quantum disordered phase.
In that phase, CHN obtain $\xi\sim
\frac{S\,a}{S_{c}-S}$
for the correlation length. It is very likely that
instanton contributions
to the Hopf term will stabilize spin-Peierls order
here \cite{read2}.
\item At finite, but not too large, temperatures in the region where
the ground state is Neel ordered ($g<g_{c}$) we can integrate all
quantum fluctuations to
map the QNL$\sigma$M into a classical problem with a
renormalized spin stiffness and a short wavelength cutoff of order
$\hbar c/k_{B}T$.
\end{itemize}

Our goal is to analyze the stability of the Neel state when
introducing frustration. We will therefore use CHN approach in the
zero temperature formalism. To begin with, we recalculate some of the
magnitudes already discussed by the former authors. We defer to
appendix C the derivation of the scaling equations, and quote here
only the results. We obtain:
\begin{eqnarray}
\frac{d\bar{g}}{dl}&=&-\bar{g}+\bar{g}^{2}\\\nonumber
\frac{d\rho}{dl}&=&-\rho \bar{g}\\\nonumber
\frac{d\sigma}{dl}&=&\sigma \bar{g}\\\nonumber\\\nonumber
\bar{g}_{0}&=&\frac{S_{c}}{S}=\frac{0.225}{S}
\end{eqnarray}
where $\sigma$ is the component of $\hat{n}$ along the quantization
axis and $\bar{g}=\frac{g}{4\pi}$. Integrating
\begin{eqnarray}
\bar{g}_{l}&=&
\frac{\bar{g}_{0}e^{-l}}{1-\bar{g}_{0}(1-e^{-l})}\\\nonumber
\rho_{l}&=&\rho_{0}(1-\bar{g}_{0}(1-e^{-l}))\\\nonumber
\sigma_{0}&=&\sigma_{\infty}(1-\bar{g}_{0})
\end{eqnarray}

We plot the flow diagram in figure 4(a).  There are two different
phases, separated by the  straight line $g=1$. To the left of this
separatrix ($g<1$),
we have the renormalized classical phase, where there is
Neel long range order renormalized by quantum fluctuations.  All the
flow lines end at the $\rho$ axis, which is a line of fixed points.
The solution of the scaling equations is:
\begin{eqnarray}
g_{\infty}&=&0\\\nonumber
\rho_{\infty}&=&
\rho_{0}(1-g_{0})=\rho_{0}(1-\frac{S_{c}}{S})\\\nonumber
<\sigma>_{0}&=&1-g_{0}=1-\frac{S}{S_{c}}\;\;S<\sigma>_{0}=S-S_{c}
\end{eqnarray}

At the right of the separatrix, we have the quantum disordered phase,
where interaction between magnons is so strong that confine them,
generating an internal length scale
$l^{*}=\ln(\frac{g_{0}}{g_{0}-1})$ in terms of which we define the
correlation length
\begin{equation}
\xi\sim ae^{l^{*}}=\frac{g_{0}a}{g_{0}-1}=\frac{S_{c}a}{S_{c}-S}
\begin{array}{ll}
\stackrel{S\rightarrow  0}{\rightarrow}&a\\
\stackrel{S\rightarrow S_{c}^{-}}{\rightarrow}&\infty\end{array}
\end{equation}
The coupling constant g diverges
and the spin stiffness goes to 0 at that length scale.

Finally, we think it
is worth remarking that although figure 4(a) shows
flow lines for $\rho\sim 0$, that result is spurious, because  in that
case the action for  the QNL$\sigma$M
is empty and one has to take  into account
quartic terms in the gradient expansion of the Heisenberg Hamiltonian.

\subsection{$J_{1}-J_{2}-J_{3}\,\,$ Model}
We are now in
a position to understand better the strongly frustrated case.
We will use a modified $\sigma$ model appropriate for both the weakly
and strongly frustrated magnets in the Neel state.
We will try to discern if the coupling constant
flows towards strong coupling along the {\em QCP}, signaling that
the magnet is disordered for whatever the value of the spin.

Close to the {\em QCP} the bare spin stiffness, $\rho(S)$,
is very small and
we are forced
to take into account effects due to quartic terms, just
as we did in our calculation of the staggered magnetization in SWT.
The action is
\widetext
\begin{eqnarray}
{\cal S} &=&
\frac{1}{2}\int_{-\infty}^{\infty}\int_{a}^{L} \,dt d^{2}r
\,
\left \{ \chi^{0}
(\partial_{t} \hat{n})^{2}-
\rho^{0}(S)
((\partial_{x}\hat{n})^{2})+(\partial_{y}\hat{n})^{2}))\right.
\\\nonumber& &
\left.+\frac{\sigma_{1}^{0}a^{2}}{12}((\partial_{xx}\hat{n})^{2}+
(\partial_{yy}\hat{n})^{2})
-\sigma_{2}^{0}a^2(\partial_{xx}\hat{n})(\partial_{yy}\hat{n})
\right \}\label{sigma}
\end{eqnarray}
\narrowtext
We take $\rho^{0}(S)$ as given by SWT (i.e.: the
stiffness renormalized by ``short'' wavelength quantum fluctuations),
instead of $\rho_{0}=
(J_{1}-2J_{2}-4J_{3})\,S^{2}$,
and use the values obtained in the gradient  expansion for  the rest
of the parameters entering in the action:
\begin{eqnarray*}
\sigma_{1}^{0}&\sim&(J_{1}-2J_{2}-16J_{3}) S^{2}\\
\sigma_{2}^{0}&\sim&J_{2} S^{2}\\
g^{0} &=& \frac{1}{\sqrt{\rho(S) \chi}a}
= \sqrt{\frac{8 J_{1}}{J_{1}-2J_{2}-4J_{3}}}
\,\frac{1}{S}
\end{eqnarray*}

We can already see one   of the problems which arise in the frustrated
case: the coupling constant depends not only on S (as in the
unfrustrated case), but also on the exchange interactions
$J_{1},\,J_{2}$ and $J_{3}$. Therefore, we  don't have such a clear
indicator about the magnetic order of the microscopic model
as before.
Moreover, $g_{0}$ diverges at the {\em QCP},
and therefore is not well-suited
for our analysis. This means that we will have to cook a
charge which doesn't contain $\rho(S)$ in the denominator.

After performing the perturbative
RG expansion we arrive at the following scaling equations:
\begin{eqnarray}
\frac{dg}{dl}&=&-g+g^{2} I\,
(1+\frac{\sigma_{1}-2\sigma_{2}}{8\rho})\\\nonumber
\frac{d\rho}{dl}&=&-\rho g I\,
(1+\frac{\sigma_{1}-2\sigma_{2}}{4\rho})\\\nonumber
\frac{d\sigma_{i}}{dl}&=&-\sigma_{i}(2+gI)\\\nonumber
\frac{dc}{dl}&=&-cgI\,\frac{(\sigma_{1}-2\sigma_{2})}{8\rho}
\end{eqnarray}

where I is the loop integral
\begin{equation}
I=\frac{\sqrt{6\rho}}
{\pi^{2}}\int_{0}^{\pi/2}\frac{d\theta}{\sqrt{24\rho-
2\sigma_{1}+(\sigma_{1}+6\sigma_{2})\sin^{2}\theta}}
\end{equation}

This integral diverges at the Lifshitz point, signaling that
the scaling analysis breaks down there. We see that this point
is quite pathological, and has to be treated in a different way from
the rest of the phase diagram. We have written down the scaling
equation for {\em c} to show that in the frustrated case, it does get
renormalized downwards. These scaling equations reduce to those
obtained by CHN for the unfrustrated case and to those of Ioffe and
Larkin when $\rho(S)=0$.

We will pass on now to the task of finding  an
unambiguous indicator of the ordering of the system. We have two
possibilities to obtain a charge which roughly go like $1/S$ and which
give us the separatrix between the ordered and disordered states:
\begin{enumerate}
\item We can choose ${\cal G}_{0}\sim \frac{1}{\sqrt{\rho}}$. This
kind of  charges
diverge if $\rho_{0}(S)$
is close to the {\em QCP}. They have a typical
behavior ${\cal G}\sim -\alpha(\rho) \,{\cal G}+\beta(\rho)\,{\cal
G}^{2}$, which means that there is a separatrix but, because ${\cal
G}_{0}>> 1$, one is always in the disordered phase.
\item ${\cal G}_{0}\sim \frac{1}{\sqrt{\sigma_{1}\sigma_{2}}}$. In
this case, the system always flows towards strong coupling and there
is no separatrix.
\end{enumerate}

There is a third possibility: we  can
evaluate the action  strictly along the {\em QCP} and define, following
Ioffe and Larkin \cite{ioffe}, the following charge:
\widetext
\begin{equation}
G^{0} = \frac{I}{\sqrt{\rho(S)} \chi a}
=\frac{A}{\chi^{1/2}\sigma_{1}^{1/4} \sigma_{2}^{1/4}a}
\sim \sqrt{\frac{J_{1}}{(J_{2} J_{3})^{1/2}}}\,\frac{1}{S} \,\,\,\,
(\rho^{0}=0)
\end{equation}
\narrowtext

We then assume that
$A= (\sigma_{1}\sigma_{2})^{1/4}I/\rho^{1/2}(S)$
doesn't get renormalized.
We have checked that the {\em QCP} is stable under iteration of
the scaling equations. We find that G satisfies:
\begin{equation}
\frac{dG}{dl}=G^{2}\;\;:G_{l}=\frac{G_{0}}{1-G_{0} (l-l_{0})}
\end{equation}
It flows towards strong coupling, which means that
it stays effectively at its lower critical dimension, with the
generation of a
correlation length $\xi \sim a e^{\frac{1}{G_{0}}}\sim a e^{S}$.

We should  draw now the separatrix between the ordered state and the
spin liquid.
Ioffe and Larkin argue that for finite $\rho(S)$
the relevant action is not given by equation 14, but by
\begin{eqnarray}
{\cal S}
&=& \frac{1}{2}
\int_{-\infty}^{\infty}\int_{\ln(\frac{1}{\rho(S)})}^{L}
\,dt d^{2}r\,
\left \{ \chi^{0}
(\partial_{t} \hat{n})^{2}\right.\\\nonumber
&&\left.+\frac{\sigma_{1}^{0}a^{2}}{12}((\partial_{xx}\hat{n})^{2}+
(\partial_{yy}\hat{n})^{2})
-\sigma_{2}^{0}a^2(\partial_{xx}\hat{n})(\partial_{yy}\hat{n})
\right \}
\end{eqnarray}

where $\ln(\frac{1}{\rho(S)})$ acts as a cutoff in the scaling
equations. In that case, the separatrix is given by
\begin{equation}
1+G_{0}\ln(\rho(S))=0
\end{equation}

We stress that Ioffe and Larkin's action is different from the naive
hydrodynamic one, equation 14. It is not clear to us
which one of them really
describes the physics of the $J_{1}-J_{2}-J_{3}\,\,$ model
close to the {\em QCP}. In any case, there is no doubt that the
separatrix has to exist, be it an exponential or not. We show in
figure 4(b) the flow
diagram of the complete $J_{1}-J_{2}-J_{3}\,\,$ model, where we draw
a tentative
separatrix for small $\rho$ and join it with the one given
for  the weakly
frustrated case. It is amusing to note that turning the
figure by $90^{0}$, the
diagram is the same as the one given by SWT and
SBMFT, if we take into account that we are writing the dressed
$\rho(S)$ in the ordinate axis.
\subsection{Discussion}
We recapitulate here the main conclusions we have reached:
\begin{itemize}
\item We have found that the system is at its lower
critical dimension along the {\em QCP}:
the charge flows to strong coupling, signaling a spin liquid state.
\item There is a separatrix between the spin liquid phase and
the Neel ordered state. That line begins at the point
$(\rho(S),g)=(0,0)$ and links the CHN separatrix for sufficient large
values of the spin stiffness.
\item The flow diagram looks very much like the stability diagram that
we obtained from SWT and SBMFT, with an enhancement of the Neel phase
due to {\em order from disorder}.
\item There is a dimensional crossover from the region $\rho(S)<<1$,
where the system is 2+1 dimensional, and $g'=-g+g^{2}$ to the region
where the spin stiffness is very small, and
the system is at its critical dimension, $d_{c}$, so that $G=G^{2}$.
\end{itemize}

\section{$J_{1}-J_{2}\,\,$ model}
We devote this section
to study the $J_{1}-J_{2}\,\,$ model. We have already argued
that it is the most complicated case of the whole phase diagram, and
probably, the least relevant. We nevertheless study it  because
it has attracted a lot of attention.

We begin, as in
section II, by plotting the phase diagram obtained from
LSWT and SBMFT (figure 5(a)).
The expressions for the magnetization of the
Neel and collinear states from LSWT as a function of
$J_{2}/J_{1}$ are \cite{chandra}:
\widetext
\begin{eqnarray}
<S^{z}>
&=&S+\frac{1}{2}-\frac{1}{N}\sum_{k}\frac{1-J_{2}(1-\Gamma_{k})}
{\sqrt{(1-J_{2}(1-\Gamma_{k}))^{2}-
\gamma_{k}^{2}}}\;\;\;\;\;\; (Neel)\\\nonumber
&=&S+\frac{1}{2}-\frac{1}{N}\sum\frac{1+\lambda\cos(k_{y})}
{\sqrt{(1+\lambda\cos(k_{y}))^{2}
-\cos(k_{x})^{2}(\lambda+\cos(k_{y}))^{2}}}
\;\;(col)
\end{eqnarray}
\narrowtext

where $\lambda=1/2J_{2}$.
The denominators are
the velocities of the spin waves. The value of S
for which the
magnetization $<S^{z}>$ goes to zero defines the boundary we
are looking for.

As already stated by other authors \cite{xu,mila},
LSWT predicts a finite region where the ground state is disordered
around the classical
tricritical point $J_{2}=0.5$ for any S (and in particular, for
$S=\frac{1}{2}$). On the contrary, mean field theories predict
that the stability of the Neel state is enhanced by quantum
fluctuations with a reentrant boundary joining the classical
frustrated point with zero slope. This fact means that there is,
ineludibly, a region of
coexistence of Neel and collinear states and therefore, a finite
hysteresis and a first order phase transition.
This first order phase transition extends even for  $S<1/2$, so that
there isn't  a spin liquid state for spin one-half.

We would like to
devote some words to some interesting features of the
SBMFT boundary of the collinear state.
In fig. 5(b) we show the
SBMFT results in a logarithmic scale. It shows a bump where
the stability of the collinear state is even bigger than that of the
unfrustrated case and then tends to the correct result for $
J_{2} \rightarrow \infty$ ($1/S\sim 5.1$). In the same figure we show
the results
obtained in numerical diagonalizations \cite{dagotto} for spin
one-half. We see that
the boundaries of the ordered states from mean field theory
are much higher than
the results of numerical diagonalizations. These two
pieces of evidence support our belief that mean field theories
overestimate the stability of the ordered states and
that there is a spin liquid in a very small region at spin $1/2$ for
$J_{2}\sim0.6$.

We  solve now the discrepancy between LSWT and SBMFT.
We evaluate  the
staggered magnetization of the Neel state in the next to
LSWT order in $1/S$ close to the {\em CCL}:
\begin{equation}
<S^{z}>\simeq
S+\frac{1}{2}-\alpha \ln^{2}(\rho_{cl})+\frac{\beta}{S}
\frac{\ln^{2}(\rho_{cl})}{\rho_{cl}}+O(\frac{1}{S^{2}})
\end{equation}

where
we use $\rho_{cl}=J_{1}-2J_{2}$
as a cutoff for the infrared divergent
integrals and $\alpha$ and $\beta$ are constants.
This series can be resumed by computing ${\cal E}(k)$, the
spin wave energy, to
next to LSWT order in $1/S$. Using the spin wave velocity
\begin{displaymath}
\rho(S)
\sim \rho_{cl}
-\frac{2J_{2}}{S}(\alpha_{2}-\alpha_{1})
\end{displaymath}
as a cutoff, instead of the spin stiffness, we find:
\begin{equation}
<S^{z}>\simeq S+\frac{1}{2}-\gamma\ln^{2}(\rho(S))
\end{equation}
which means that $1/S$ corrections to LSWT just renormalize the CCL
where the spin stiffness (and spin wave velocity) is zero.
The {\em QCP} for the
Neel state along which the spin stiffness is zero
gets tilted to the
right as one goes further in perturbation theory in
$1/S$. Therefore, SWT
also predicts an enhancement of the stability of
the Neel state and therefore, {\em order from disorder}.
In other words: it tends to coincide with SBMFT.

An
interesting feature
of equations 22 and 23 is that $<S^{z}>$ diverges as
the square of the logarithm of the cutoff, signaling that the
Renormalization Group analysis does not work at that point.

\section{Conclusion}
The purpose of this paper was to shed some new light on
the recent controversy about
the existence of a spin
liquid phase in the phase diagram of the $J_{1}-J_{2}\,\,$
and $J_{1}-J_{2}-J_{3}\,\,$
models. We have tried to do a thorough analysis of the
problem using different techniques, comparing them and extracting
consequences about the validity of each one. We have obtained a very
consistent qualitative picture of how the phase diagram looks like.
For the $J_{1}-J_{2}\,\,$ model,
there seems to be a first order phase transition,
due to the fact that the stability of the Neel state is enhanced by
quantum fluctuations.
At the same time there probably
exists a very small region where the magnet is
disordered for spin $\frac{1}{2}$. Its frustrated point, though, is a
Lifshitz point and therefore is not representative of the phase
diagram of the $J_{1}-J_{2}-J_{3}\,\,$ model.
In the $J_{1}-J_{2}-J_{3}\,\,$ model we have found {\em order from
disorder}, too. But now there is a finite region
where the system is disordered for any value of the spin, because the
stability of the helicoidal phase is reduced as much as the one of the
Neel phase is enhanced, and there is no first order phase transition.
For spin $\frac{1}{2}$ the spin liquid region
is shifted from the {\em Classical Critical line}
$J_{1}-2J_{2}-4J_{3}=0$ (figure 1) to
considerably bigger values of $J_{2}$ and $J_{3}$.
We find that the character of the spin disordered phase seems to be
different for the weakly  and for the strongly frustrated cases. This
fact could have any repercussion in the role of instanton tunneling
between different topology sectors, and the stabilization of
spin-Peierls order close to the critical line.
Finally, we would like to mention that our results agree with
numerical diagonalization studies by Dagotto et al
\cite{moreo,dagotto} both for
the $J_{1}-J_{2}-J_{3}\,\,$ and the $J_{1}-J_{2}\,\,$ models
but are in contradiction with the recent work of Schulz and Ziman
\cite{schulz} and with the results obtained from series expansions
\cite{series} for the $J_{1}-J_{2}\,\,$ model.

\acknowledgments
We would like to thank P. Coleman, L. Ioffe, F. Gebhard,
N. Andrei, H. Neuberger, M. B\`{e}zard, X. Y. Zhang,
P. Chandra and F. Mila for illuminating discussions.
The financial support from the Fulbright-MEC (Spain) program is also
gratefully acknowledged.

\appendix{Spin Wave Theory for an helical magnet}
For a general Heisenberg model
\begin{equation}
{\cal H}= \sum_{ij} J_{ij} \vec{S}_{i}\cdot\vec{S}_{j}
\end{equation}
a classical helical configuration
$\vec{S}_{i}=S(\hat{u}\sin(\vec{Q}\cdot\vec{R}_{i})
+\hat{v}\cos(\vec{Q}\cdot\vec{R}_{i}))$,
with $\hat{u}\cdot\hat{v}=0$,
has an energy $E_{cl}=N\,S^{2}
J(\vec{Q})$.
This helical state lies in the plane normal to the ``twist''
vector $\hat{n}=\hat{u}\times\hat{v}$.
The classical ground state is the helix with
pitch $\vec{Q}_{0,cl}$ obtained
by minimizing $E_{cl}$, i.e.: $\vec\partial_{\vec{Q}}
J(\vec{Q}_{0,cl})=0$.
To perform a spin wave expansion about this classical
state we
rotate the local quantization
axis through an angle $\theta_{i}=\vec{Q}\cdot\vec{R}_{i}$
about the twist vector $\hat{n}$,
transforming to a twisted reference
frame in which the ordered state is ferromagnetically aligned
\cite{rastelli,ritchie}:
\begin{displaymath}
\vec{S}_{i} \rightarrow
e^{\vec{A}_{i}\times}\vec{S}_{i}=
e^{\vec{Q}\cdot\vec{R}_{i}\hat{n}\times}\vec{S}_{i}
\end{displaymath}
with this prescription, we obtain the following Hamiltonian:
\widetext
\begin{equation}
{\cal H}'= \sum_{ij}J_{ij}
\left\{\gamma_{ij}^{+}\vec{S}_{i}\cdot\vec{S}_{j}+\gamma_{ij}^{-}
(2\,\,\vec{S}_{i}\cdot\hat{n}\,\,\vec{S}_{j}\cdot
\hat{n}-\vec{S}_{i}\cdot\vec{S}_{j})
-\sin(\vec{Q}\cdot\vec{R}_{ij})
\hat{n}\cdot(\vec{S}_{i}\times\vec{S}_{j})
\right\}
\end{equation}
\narrowtext
where $\gamma_{ij} =
\frac{1}{2}\,(1\pm\cos(\vec{Q}\cdot\vec{R}_{ij}))$.
We choose $\hat{n}=\hat{z}$
to carry out the calculations of this section.

The usual way to deal with these magnetic models is to decompose
the spins into Holstein-Primakoff (HP) bosons \cite{mattis,dyson}
and treat the resulting Hamiltonian as a dilute gas of bosons
\cite{harris}.
We  use this representation to calculate LSWT properties
of the different spin configurations we are dealing with.
The ground state energy of the helicoidal $\vec{Q}=(q,q)$ state is
\widetext
\begin{eqnarray}
\frac{E_{0}}{N}&=&
\frac{S(S+1)}{2}\,J_{Q_{0}}+\frac{S}{2}\sum_{k}E(k)\\\nonumber
E(k)&=&\sqrt{(J_{Q_{0}}-J_{k})(J_{Q_{0}}-\frac{1}{2}
(J_{k+Q_{0}}+J_{k-Q_{0}}))}=
\sqrt{(J_{Q_{0}}-J_{k})(J_{Q_{0}}-J_{\pm})}
\end{eqnarray}
\narrowtext

where again $\vec{Q_{0}}$
is determined by minimizing $E_{0}$. We can
expand $\vec{Q}_{0}$ in powers of 1/S about its classical value
($q_{0,cl}=-\frac{1}{2J_{2}+4J_{3}}$ for the state
$\vec{Q_{0}}=(q_{0},q_{0})$). The result is
\widetext
\begin{equation}
q_{0}^{i} - q_{0,cl}^{i}=\Delta q_{0}^{i}=
\frac{1}{2 S \partial_{ii}J_{Q}}\sum_{k}\partial_{i}
J_{\pm}
\left(\frac{J_{k}-J_{Q}}{J_{\pm}-J_{Q}}\right)^{1/2}
+O(\frac{1}{S^{2}})
\end{equation}
\narrowtext

The spin stiffness is defined as the second derivative of the free
energy with respect to a uniform twist to the direction of the order
parameter \cite{rudnick}. This is equivalent to the response function
to a gauge field $\vec{A}_{i}=\delta{\vec{Q}}\cdot\vec{R}_{i}\hat{k}$
applied to the system, where $\hat{k}$ is the axis of the twist and
$\delta \vec{Q}$, its wavevector. Then,
\begin{equation}
\rho^{a}_{ij}=\frac{\partial^{2}F}{\partial A^{a}_{i}\partial
A^{a}_{j}}
\end{equation}
For an helical magnet, with $\hat{n}\parallel\hat{z}$, we have
$\rho^{x}_{ii}= \rho^{y}_{ii}\neq\rho^{z}_{ii}$ and
$\rho^{a}_{xx}=\rho^{a}_{yy}$.
For the purpose of this paper, it will
be enough to calculate $\rho^{z}_{ii}$. We obtain:
\widetext
\begin{eqnarray}
\rho_{ii}^{z}&=&\frac{S^{2}\partial_{ii}J_{Q_{0,cl}}}{2}\\\nonumber
&&+\frac{S}{2}\left\{
\partial_{ii}J_{q_{0,cl}}-(\sin(q_{0,cl})+8
J_{3}\sin(2Q_{cl}))S\Delta q_{0}^{i}\right.\\\nonumber&&\left.
+\frac{1}{2}\sum_{k}\frac{(2J_{Q_{0,cl}}-J_{k}-J_{\pm})
\partial_{ii}J_{Q_{0,cl}}-(J_{Q_{0,cl}}-J_{k})\partial_{ii}
J_{\pm}}{E(k)}
\right\}+O(1)
\end{eqnarray}
\narrowtext

On the other hand, for a bipartite lattice, and when the ground state
is Neel ordered $(\pi,\pi)$, it proves more advantageous to  use a
Dyson-Maleev
(DM) decomposition of the spins
\cite{mattis}, because then, the Hamiltonian
doesn't have terms higher than fourth order and because corrections
to mean field results are less important than the ones obtained with
HP bosons. The results obtained with this method are
identical to those of HP at the mean field level as we
will show \cite{harris}.

We want to  calculate the staggered magnetization for the
Neel state of the two  models
to next to LSWT order in $1/S$ using DM to check the convergence
of the series. The Hamiltonian is given by equation A2, with $\vec{Q}=
(\pi,\pi)$. We perform a
Fourier transform after introducing DM bosons and rewrite it as:
\widetext
\begin{eqnarray}
{\cal H} = &&S z \sum_{k}\left\{A(k)\,(a_{k}^{+}a_{k}+b_{k}^{+}b_{k})
+ J_{1}\gamma_{k}(a_{k}b_{-k}+a_{k}^{+}b_{-k}^{+})\right\}\\\nonumber
&-& \frac{zJ_{1}}{N}\sum_{1234}\delta(1+2-3-4)\left\{
2\,\gamma_{1+2} a_{1}^{+}a_{-2}b_{-3}^{+}b_{4}
+\gamma_{1} a_{1}^{+}b_{2}^{+}b_{-3}^{+}b_{4}
+\gamma_{4} a_{1}^{+}a_{-2}^{+}a_{3}b_{4}\right\}\\\nonumber
&+& \frac{zJ_{\delta}}{N}\sum_{1234}\delta(1+2-3-4)\left\{
(\Gamma^{\delta}_{2}-
\Gamma^{\delta}_{2-4}) a_{1}^{+}a_{2}^{+}a_{3}^{+}a_{4}
+(\Gamma^{\delta}_{4}-
\Gamma^{\delta}_{2-4}) b_{1}^{+}b_{2}^{+}b_{3}b_{4}\right\}
\end{eqnarray}
\narrowtext

where
\begin{eqnarray*}
A(k)&=&J_{1}-J_{\delta}(1-\Gamma^{\delta}_{k})\\\nonumber
\gamma_{k}&=&\frac{1}{2}(\cos(k_{x}a)+\cos(k_{y}a))\\\nonumber
\Gamma^{2}_{k}&=&\cos(k_{x}a)\cos(k_{y}a)\\\nonumber
\Gamma^{3}_{k}&=&\gamma_{2k}
\end{eqnarray*}
and z is the coordination number.

We introduce a Bogoliubov transformation,
$a_{k}^{+}=u\alpha_{k}^{+}+ v \beta_{-k}$
and $b_{-k}=v\alpha_{k}^{+}+u \beta_{-k}$, and normal-order
all terms of the Hamiltonian
with respect to the vacuum of $\alpha$ and
$\beta$, neglect the contribution coming from
normal-ordered terms involving four Bogoliubov bosons and end up with
a quadratic diagonal Hamiltonian. The procedure is standard
\cite{fetter}.
We obtain a self-consistent equation for the Bogoliubov
amplitude ${\cal V}_{k}$, by requiring that the terms proportional to
$\alpha^{+}\beta^{+}$ and $\alpha\beta$ are zero:
\widetext
\begin{eqnarray}
{\cal V}(k)&=&-\frac{sgn(F_{2}(k))}{\sqrt{2}}
\left\{\frac{F_{1}(k)}{{\cal E}(k)}-1\right\}^{1/2}\\\nonumber
F_{1}(k)&=&A(k) S-\frac{2}{N}\sum_{k'}\left\{{\cal V}^{2}(k')(1-
J_{2} (1-\Gamma^{\delta}_{k}-
\Gamma^{\delta}_{k'}+\Gamma^{\delta}_{k-k'}))+\gamma_{k'}
{\cal U}(k'){\cal V}(k')\right\}\\\nonumber
F_{2}(k)&=&\gamma_{k} S+
\frac{2}{N}\sum_{k'}\left\{\gamma_{k}
{\cal V}(k')^{2}+
\gamma_{k+k'}{\cal U}(k'){\cal V}(k')\right\}\\\nonumber
{\cal E}(k)&=&\sqrt{F_{1}^{2}(k)-F_{2}^{2}(k)}
\end{eqnarray}
\narrowtext

here ${\cal E}(k)$ is the energy of the spin waves and we have used
units where $J_{1}=1$. This expressions
can be expanded in powers of $1/S$. Doing so  we obtain the following
relation for the staggered magnetization  in terms of LSWT functions:
\widetext
\begin{equation}
<S^{z}>\label{magnetization}
\simeq S+\frac{1}{2}-\frac{1}{N}\sum_{k}\frac{A(k)}{E(k)}-
\frac{J_{\delta}}{N^{2} S}
\sum_{k}\frac{\gamma_{k}^{2}
(1-\Gamma^{\delta}_{k})}{E^{3}(k)}
\sum_{k'}
\frac{A(k')\Gamma^{\delta}_{k'}-\gamma_{k'}^{2}}{E(k')}
\end{equation}
\narrowtext

where
\begin{displaymath}
v(k)^{2}= \frac{1}{2}\left\{\frac{A(k)}{E(k)}-1\right\}\;\;\;\;\;\;
E(k)= \sqrt{A^{2}(k)-\gamma^{2}_{k}}
\end{displaymath}
$E(k)$ is the energy of the spin waves in LSWT.
We note here that equation~\ref{magnetization} agrees with the
expression for the staggered magnetization obtained in reference 2.

We  can  also compute ${\cal E}(k)$ to
next to LSWT order in $1/S$ in the long wavelength limit in terms of
the spin wave velocity
\widetext
\begin{eqnarray}
c^{2}(S) &\simeq&8J_{1}S\,a\left\{
\rho_{0}\left(1+\frac{2}{S}(1+\alpha_{1}-\alpha_{4})
\right)
-\frac{2J_{2}}{S}(\alpha_{2}-\alpha_{1})
\right\}\;\;\;\;(J_{3}=0)\\\nonumber
&\simeq&8J_{1}\,S\,a
\left\{\rho_{0}\left(1+\frac{2}{S}(1+\alpha_{1}-\alpha_{4})
\right)
-\frac{4J_{3}}{S}
(\alpha_{3}-\alpha_{1})\right\}\;\;\;\;(J_{2}=0)\\\nonumber
\alpha_{1}&=&\sum_{k}\frac{\gamma_{k}^{2}}{E(k)}\;\;\;\;\;\;\;\;\;\;
\alpha_{2}=\sum_{k}\frac{A(k)\Gamma^{2}_{k}}{E(k)}\\\nonumber
\alpha_{3}&=&
\sum_{k}\frac{A(k)\Gamma^{3}_{k}}{E(k)}\;\;\;\;\;\;\;\;\;\;
\alpha_{4}=\sum_{k}\frac{A(k)}{E(k)}\nonumber
\end{eqnarray}
\narrowtext

\appendix{Selfconsistent
equations for Schwinger Bosons Mean Field Theory}
Some years ago,
Arovas and Auerbach \cite{arovas} introduced a $SU(N)$
functional integral expansion as a useful tool to deal with
unfrustrated magnets which are disordered due to quantum or thermal
fluctuations. It was based on a $SU(N)$ generalization of the spin
decomposition
\begin{displaymath}
\vec{S}_{i}=\frac{1}{2}{\bf b}_{i}^{+}\vec{\sigma}
{\bf b}_{i}\;\;\;
{\bf b}_{i}^{+}{\bf b}_{i}=2S\;\;\;
{\bf b}_{i}^{+}=(b_{i\uparrow}^{+},b_{i\downarrow}^{+})
\end{displaymath}
originally due to Schwinger \cite{mattis}. We will use here
the generalization to frustrated magnets due to Coleman and coworkers
\cite{piers,mila} for $SU(2)$.

We begin
by rotating the local quantization axis an angle $\theta_{i}=
\vec{Q} \cdot\vec{R}_{i}$ so that in the twisted reference frame the
spins are ferromagnetically aligned,
and introduce the
triplet Cooper and singlet particle-hole even-parity fields
\begin{displaymath}
\vec{B}^{+}_{ij}=
i{\bf b}^{+}_{i}\vec{\sigma}{\bf b}^{+}_{j}\;\;\;\;\;\;
D^{+}_{ij}={\bf b}^{+}_{i}{\bf b}_{j}
\end{displaymath}
with which, the action reads (choosing $\hat{n}=\hat{z}$):
\widetext
\begin{eqnarray}
{\cal S}&=&
\int_{0}^{\beta} d\tau\left\{{\bf b}^{+}\partial_{\tau}{\bf
b}-2SN\lambda-\frac{NS^{2}J_{Q}}{2}+\lambda D+\frac{1}{4}
DJ^{+}D-\frac{1}{4}B^{z,+}J^{-}B^{z}\right\}\\\nonumber
J^{\pm}_{kk'}&=&\frac{1}{2}\left\{J_{k-k'}\pm\frac{1}{2}
(J_{k-k'+Q}+J_{k-k'-Q})\right\}
\end{eqnarray}
\narrowtext
where summation over k,k' is understood, and $\lambda_{i}(\tau)$ is a
Lagrange multiplier which enforces the constraint at all sites and
times, and which can be regarded as another field.

We perform a Hubbard-Stratonovich
transformation from $D_{k}$, $B_{k}$ to the bond fields
$h_{k}$, $\Delta_{k}$ and make a
saddle-point approximation of the resulting action in those bond
fields. At the saddle-point, they are time independent, spatially
uniform and real (and so is the field $\lambda$).
After diagonalizing the resulting mean field Hamiltonian with
a Bogoliubov transformation
$b_{k\sigma}=u\gamma_{k\sigma}+v\gamma^{+}_{-k-\sigma}$
we find the
following Free Energy
\widetext
\begin{eqnarray}
\frac{F}{N}&=&-(2S+1)\lambda -\frac{S^{2}J_{Q}}{2}-h_{k}
J^{+,-1}_{kk'}h_{k'} +\Delta_{k}
J^{-,-1}_{kk'}\Delta_{k'}+\frac{2}{\beta}\ln\left(2
\sinh \frac{\beta \epsilon(k)}{2}\right)\\\nonumber
\epsilon(k)&=&\sqrt{(h(k)+\lambda)^{2}-\Delta^{2}(k)}
\end{eqnarray}
\narrowtext

When the ground
state is ordered, we have to allow for a one particle
condensate
contribution to the
mean field equations \cite{nozieres,sarker,mila}.
Besides, the gap in the spin wave spectrum closes --we have
a Goldstone mode--,
so that we have to adjust $\lambda$ to fulfill this
requirement. The saddle point of F at $T=0$ is given by the following
equations:
\widetext
\begin{eqnarray}
\frac{F}{N}&=&-(2S+1)\lambda-\frac{S^{2}J_{Q}}{2}-
\sum_{kk'}\alpha_{k}J_{kk'}^{+}\alpha_{k'}+
\sum_{kk'}\beta_{k}J_{kk'}^{-}\beta_{k'}\\\nonumber
S+\frac{1}{2}&=&S^{*}\delta_{k0}+\sum_{k}\frac{h_{k}+
\lambda}{2\epsilon(k)}\\\nonumber
\alpha_{k}&=&
S^{*}\delta_{k0}+\frac{h_{k}+\lambda}{2\epsilon(k)}\\\nonumber
\beta_{k}&=&
S^{*}\delta_{k0}+\frac{\Delta_{k}}{2\epsilon(k)}\\\nonumber
\epsilon(k)&=&\sqrt{(h_{k}-\lambda)^{2}-\Delta^{2}(k)}
\end{eqnarray}
\narrowtext

In general, the set of values of the pitch $\vec{Q}$ for the ground
state are obtained by minimizing F with respect to $\vec{Q}$. We can
obtain F as a series in $1/S$ by iterating the mean field equations.
We find
\begin{eqnarray}
\epsilon(k)&=&S\sqrt{(J_{Q}-J_{k})
(J_{Q}-\frac{1}{2}(J_{k+Q}+J_{k-Q}))}\\\nonumber
\frac{F}{N}&=&\frac{S^{2}J_{Q}}{2}+SJ_{Q}+\sum_{k}
\epsilon(k)=F_{0}\,S^{2}+F_{1}\,S
\end{eqnarray}
The expression for the
spin wave energy coincides with that given by
LSWT while
$F_{1}$ is twice the value given by LSWT (as already stated by
Arovas and Auerbach \cite{arovas}
for the unfrustrated case, SBMFT overcounts the
number of degrees of freedom).

SBMFT has two major problems in dealing with helicoidal states:
(1) Symmetry considerations require the
existence of Goldstone bosons at $(0,0)$ and $(q,q)$. We can adjust
$\lambda$ so that we have one of them, say $(0,0)$. But then, we don't
have the other one: $|h_{Q}+\Delta_{0}-h_{0}|\neq|\Delta_{Q}|$
(for Neel and collinear states we do get both zero modes).
This leads to unphysical results for the
stability diagram of
the helicoidal phase, and in particular, to a first
order phase transition. (2) We have already shown that $F_{1}$ the
contribution of
order S to the free energy is twice the correct value.
This means that the pitch $\vec{Q}$ obtained in the minimization is
not the correct one.

\appendix{Derivation
of the scaling equations for the long wavelength action}
In this appendix we will use the momentum-shell method to derive
recursion relations for the S-matrix of the hydrodynamic action of the
$J_{1}-J_{2}-J_{3}\,\,$ model, equation~\ref{sigma}.
This is a generalized $O(3)/U(1)=S^{2}$ model.
The unit vector $\hat{n}=(\hat{\pi},\sigma)$; $\hat{\pi}$
is a $n-1=2$-component
vector, and $\sigma$ is the direction of quantization.
The S-matrix is given by
\begin{equation}
{\bf S}(-\infty,\infty)=\int {\cal D}\hat{n}\prod_{x,t}
\delta(\sigma^{2}
(\vec{x},t)+ \hat{\pi}^{2}(\vec{x},t)-1)e^{i{\cal S}}
\end{equation}
It is convenient
to define the dimensionless variables $\tilde{r}=r/a$ and
$\tilde{t}=c\,t/a$ so that the action is
\widetext
\begin{eqnarray}
{\cal S} &=& \frac{1}
{2g_{0}}\int_{-\infty}^{\infty}\int_{1}^{L} \,d\tilde{t}
d^{2}\tilde{r}\, \left \{ (\partial_{\tilde{t}} \hat{n})^{2}-
((\partial_{\tilde{x}}\hat{n})^{2})+
(\partial_{\tilde{y}}\hat{n})^{2}))\right.\\\nonumber& &
\left.+\frac{\sigma_{1}^{0}}{12\rho(S)}
((\partial_{\tilde{x}\tilde{x}}\hat{n})^{2}+
(\partial_{\tilde{y}\tilde{y}}\hat{n})^{2})
-\frac{\sigma_{2}^{0}}{\rho(S)}
(\partial_{\tilde{x}\tilde{x}}\hat{n})
(\partial_{\tilde{y}\tilde{y}}\hat{n})+\frac{2HSn^{z}}{\rho(S)}
\right \}
\end{eqnarray}
\narrowtext

We integrate
$\sigma(\vec{x},t)$ now, and generate an action which is
nonlinear in the fields $\hat{\pi}(\vec{x},t)$. The nonlinear terms
constitute the interaction \cite{amit}. As we are interested in one
loop corrections to the charge g, we retain only the quartic terms.
We divide the action now into a free part and the interaction
\widetext
\begin{eqnarray}
{\cal S}_{0}&=&
\frac{1}{2g_{0}}\int\frac{dwd^{2}k}{(2\pi)^{2}}\hat{\pi}(k,w)
\cdot\hat{\pi}(-k,-w)
\\\nonumber&&\times
\left\{w^{2}-k^{2}-g_{0}h+
\frac{\sigma_{1}}{12\rho(S)}(k_{x}^{4}+k_{y}^{2})-
\frac{\sigma_{2}}{\rho(S)}k_{x}^{2}k_{y}^{2}\right\}\\\nonumber
{\cal S}_{1}&=&
\left(\prod_{i=1-4}\int\frac{dw_{i}dk^{2}_{i}}{(2\pi^{2})}\right)
\hat{\pi}(1)\cdot\hat{\pi}(2)\;\hat{\pi}(3)\cdot
\hat{\pi}(4)\delta^{3}(1+2+3+4)\\\nonumber
&&\times\left\{
w_{2}w_{4}-k_{2}
\cdot k_{4}-\frac{g_{0}h}{4}+\frac{\sigma_{1}}{12\rho(S)}
(k_{x1}k_{x2}k_{x3}k_{x4}+k_{x2}^{2}
k_{x4}^{2}+2k_{x1}k_{x2}k_{x4}^{2}+(x
\leftrightarrow y))\right.\\\nonumber&&\left.
-\frac{\sigma_{2}}{\rho(S)}(k_{x1}k_{x2}k_{y3}k_{y4}+
k_{x2}^{2}k_{y4}^{2} +2k_{x1}k_{x2}k_{y4})\right\}
+\frac{i}{2\Lambda_{t}}
\int\frac{dwd^{2}k}{(2\pi)^{2}}\hat{\pi}(k,w)\cdot\hat{\pi}(-k,-w)
\end{eqnarray}
\narrowtext

where $h=\frac{a^{3}SH}{c}$ and $\Lambda_{t}$ is an ultraviolet cutoff
for the time direction. The free
action defines an unperturbed propagator ${\cal G}_{0}$, which we use
to compute the different terms which arise in the loop expansion. We
compute the renormalization of the free term to one loop order, by
expanding $e^{S_{1}}$, contracting fields and exponentianting again.
The measure cancels
the ultraviolet divergent terms arising in the one
loop expansion as it does for the classical case. In particular, the
regulator $\Lambda_{t}$ doesn't appear in the renormalized theory.
The procedure is very well explained in reference 30, so we
refer the reader to it for more details. We obtain
\widetext
\begin{eqnarray}
{\cal S}'_{0}=&&\int\frac{dw}{2\pi}
\int_{0}^{e^{-l}}\frac{d^{2}k}{(2\pi)^{2}}
\hat{\pi}(k,w)\cdot\hat{\pi}(-k,-w)\\\nonumber
&&\times\left\{w^{2}
(1+g_{0}I)-k^{2}(1+g_{0}I(1-\frac{\sigma_{1}-2\sigma_{2}}
{4\rho(S)}))+\frac{\sigma_{1}}{12\rho(S)}(k_{x}^{4}+k_{y}^{4})(1+g_{0}I)
\right.\\\nonumber&&-\left.
\frac{\sigma_{2}}{\rho(S)}k_{x}^{2}k_{y}^{2}(1+g_{0}I)-g_{0}h(1+g_{0}I)
\right\}
\end{eqnarray}
\narrowtext

where I is the following loop integral
\widetext
\begin{eqnarray*}
I&=&\int_{-\infty}^{\infty}
\frac{dw}{2\pi}\int_{e^{-l}}^{1}\frac{d^{2}k}{(2\pi)
^{2}}\frac{i}{w^{2}-k^{2}+
\frac{\sigma_{1}}{12\rho(S)}(k_{x}^{2}+k_{y}^{2})-
\frac{\sigma_{2}}{\rho(S)}k_{x}^{2}k_{y}^{2}}\\\nonumber&=&
\frac{\sqrt{6\rho(S)}\,l}
{\pi^{2}}\int_{0}^{\pi/2}\frac{d\theta}{\sqrt
{24\rho(S)-2\sigma_{1}+(\sigma_{1}+6\sigma_{2})\sin^{2}\theta}}
\end{eqnarray*}
\narrowtext
It diverges at the tricritical point, showing that the scaling
analysis can not be
applied to dilucidate what happens in the $J_{1}-J_{2}\,\,$
model close to the tricritical point.

After performing the rescaling
\begin{eqnarray*}
k&=&e^{-l}\,k'\;\;\;\;\;\;w=\frac{c'}{c}e^{-l}w'\\\nonumber
\hat{\pi}&=&\zeta\hat{\pi}'\;\;\;\;\;\;h'=\zeta\,h
\end{eqnarray*}
we find the parameters of the renormalized action in terms of the old
ones:
\begin{eqnarray}
(g\,h)'&=&e^{2l} (g\,h)
\left(1+gI\frac{\sigma_{1}-2\sigma_{2}}{4\rho(s)}\right)\\\nonumber
(\frac{1}{g})'&=&\frac{1}{g}\zeta^{2}e^{-5l}
\left(1+gI(1-\frac{3}{8}\frac{\sigma_{1}
-2\sigma_{2}}{\rho(S)})\right)\\\nonumber
c'&=&c\left(1-gI\frac{\sigma_{1}
-2\sigma_{2}}{8\rho(S)}\right)\\\nonumber
\sigma_{i}&=&\sigma_{i}(1-2l-gI)\\\nonumber
\rho'(S)&=&\rho(S)
\left(1-gI(1+\frac{\sigma_{1}-2\sigma_{2}}{4\rho(S)})\right)
\end{eqnarray}
from which one obtains easily the scaling equations of the model after
iterating.

\figure{ Classical
Phase diagram for
the $J_{1}-J_{2}-J_{3}\,\,$ model. Thick solid lines
indicate a discontinuous transition; dashed lines indicate a
continuous one. The thin lines are the results of LSWT for the
boundaries of the different ordered states for spin 1/2.}
\figure{ Phase diagram for
the $J_{1}-J_{2}-J_{3}\,\,$ model. Dashed lines are LSWT; solid,
SBMFT. The SWT $\rho_{Neel}$ vanishes along the dotted line.  }
\figure{(a) LSWT correction to $q_{0,cl}$ as defined in equation 8.
(b) Classical spin stiffness (solid lines) and LSWT corrections to
it (dashed lines).}
\figure{(a)
Flow diagram for the QNL$\sigma$M in 2+1 dimensions. The dashed
line separates the Renormalized Classical region from the Quantum
Disordered one.
(b) Flow diagram for the $J_{1}-J_{2}-J_{3}\,\,$ model, linking
CHN and Ioffe and Larkin analysis. We plot in the y-axis $\rho(S)$,
the spin stiffness renormalized by short-wavelength quantum
fluctuations. Rotating the figure by $90^{0}$ we obtain the phase
diagram of SWT.}
\figure{ (a)
Phase diagram
for the $J_{1}-J_{2}\,\,$ model. Dashed lines are LSWT results,
while solid lines are those given by SBMFT. (b) same as before in a
logarithmic scale. The cross is the result from numerical
diagonalization of small clusters \cite{moreo}}

\end{document}